\documentclass[12pt,a4paper,showkeys]{revtex4}
\usepackage{graphicx}
\usepackage{epsfig}
\usepackage{bm}
\usepackage{amsmath}
\usepackage{amssymb}
\usepackage{graphics}
\usepackage{epstopdf}
\usepackage[linkcolor=red,citecolor=green,urlcolor=blue]{hyperref}

\begin{document}

\title{ \textbf{A second update on double parton distributions}}
\author{Federico Alberto Ceccopieri}
\email{federico.alberto.ceccopieri@cern.ch}
\affiliation{IFPA, Universit\'e de Li\`ege,  All\'ee du 6 ao\^ut, B\^at B5a,   \\4000
Li\`ege, Belgium}
\begin{abstract}
We present two equivalent consistency checks of the momentum sum rule for  double parton distributions and show the importance of the inclusion of the so-called 
inhomogeneous term in order to preserve correct longitudinal momentum correlations.
We further discuss in some detail the kinematics of the splitting at the basis 
of the inhomogeneous term and update the double parton distributions evolution equations at different virtualities.

\end{abstract}


\maketitle

\section{Introduction}
\noindent
The hadron internal structure is presently encoded, thanks to the QCD factorisation theorem, in process-independent parton distribution functions (PDFs). The latter allow to predict cross sections for high-mass systems and high transverse-momentum jets in hadronic collisions  
in terms of binary partonic interactions.
There are, however, increasing experimental evidences (for recent analyses see Ref.~\cite{DPS_exp_new}) that hard double parton scattering (DPS) may occur within the same hadronic collision. The experimental and theoretical efforts to identify and quantify DPS contributions aim to understand and control this additional QCD background in new physics
searches, especially in the multi-jet channel. At a more fundamental level, DPS could unveil parton correlations in the hadron structure not accessible in single parton scattering (SPS) and encoded in novel distributions, \textsl{i.e.} double parton distributions (DPDs).
So far, measuraments have only provided informations on $\sigma_{eff}$.
This dimensionful parameter controls the magnitudo of DPS contribution  
under the simplifying assumptions of two uncorrelated hard scatterings and full factorisation 
of DPDs in terms of ordinary PDFs and model-dependent distribution in transverse position space. Many theoretical analyses have predicted QCD evolution effects on DPDs
relaxing some or all the above assumptions~\cite{snigirev2003,BDFS_sum_rule,Ceccopieri_DPD,MW,DS}. A part recent progress in this direction reported in Ref.~\cite{D014}, the experimental observation of the expected mild scaling violations induced by DPDs evolution is not yet possible given the accuracy of the present data. Nonetheless, a good theoretical control of the latter is mandatory if the whole DPS formalism has to be properly validated against data.
A first attempt to calculate the scale dependence of longitudinal DPDs (hereafter called lDPDs) has been presented long ago
in Ref.~\cite{snigirev2003} under the assumption of factorisation in transverse space. With respect to standard single-parton distributions~\cite{DGLAP}, lDPDs evolution equations do contain an additional 
term which is responsable for perturbative longitudinal correlation between the interacting partons. This result has stimulated in the recent past an increasing activity 
in the field and has generated some constructive criticism in the literature.
A first critical point is that the relative transverse momentum 
of the interacting parton pair is not conserved between amplitude and its conjugated~\cite{BDFS_4jet}. This implies that one should consider new distributions, addressed as 
two-particle generalised parton distributions, $_{2}\mbox{GPD}s$, 
which have an additional dependence on a transverse momentum vector $\Delta$ which parametrises this imbalance~\cite{BDFS_4jet}. They reduce to lDPDs addressed in this paper when this vector is set to zero or, in position space, if they are integrated over the relative distance $b$ of the parton pair. This additional dependence affects the evolution of the 
correlated and uncorrelated terms in rather different way~\cite{DS}
and give rise to inconsistencies with respect to the formalism of Refs.~\cite{snigirev2003,Ceccopieri_DPD}. More importantly, $_{2}\mbox{GPD}s$ enter the DPS cross sections rather than their $b$-integrated or $\Delta=0$ counterparts, \textsl{i.e.} longitudinal DPDs, and moreover the integral over the imbalance $\Delta$ of the product 
of $_{2}\mbox{GPD}s$ is directly proportional to the value of $\sigma_{eff}^{-1}$~\cite{BDFS_4jet,BDFS_12}.  
A second critical point is that the inclusion of single splitting contributions, 
according to the formalism of Ref.~\cite{snigirev2003}, poses a problem 
of consistency with SPS loop corrections when DPDs are used to evaluate
DPS cross sections. A problem which is solved
if one considers two-particle generalised parton distributions, $_{2}\mbox{GPD}s$~\cite{GS11}.
From these observations, it appears that $_{2}\mbox{GPD}s$ offer a natural solution to this 
class of problems and are a good candidate to focus on when addressing 
the issues related to QCD evolution. 
On the other hand, as we shall describe in the following, the presence of the inhomogeneous term in the evolution equations appearing Ref.~\cite{snigirev2003} is crucial if one demands that longitudinal DPDs satisfie QCD consistency check for the momentum sum rule. It appears therefore that the road towards a consistent treatment of QCD evolution effects on DPDs is quite narrow as it must reconcile all these requirements at once. 

This paper is organized as follows. In Sec.~\ref{preliminar}
we collect some definitions and formulas pertinent to the Jet Calculus formalism~\cite{KUV} and frequently used thereafter. In Sec.~\ref{MSR} we present two equivalent derivations of the momentum sum rule for lDPDs, paying particular attention to some delicate steps
occurring in the calculation. In Sec.~\ref{kine} we discuss  in some detail the kinematics of the splitting in the inhomogeneous term and update the lDPDs evolution equations at different virtualities in light of the results obtained for the momentum sum rule. We summarise our results in Sec.~\ref{summary}.

\section{Preliminaries}
\label{preliminar}
\begin{figure}[t]
\includegraphics[width=13cm,height=5cm,angle=0]{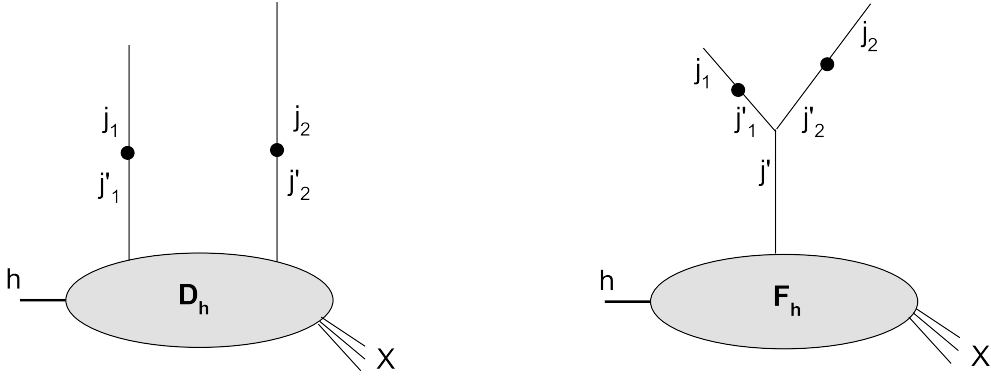}
\caption{Pictorial representation of both terms on right hand side of eq.~(\ref{Ddef}).
Black dots symbolize the parton-to-parton evolution function, $E$\,. }
\label{fig1}
\end{figure}

We recall briefly the main ingredients which we will use in our calculations.
The longitudinal double-parton distributions $D_h^{j_1,j_2}(x_1,Q_1^2,x_2,Q_2^2)$
are interpreted as the two-particle inclusive distribution to find in a target hadron a couple of partons of flavour $j_1$ and $j_2$ with fractional momenta $x_1$ and $x_2$ and
virtualities up to $Q_1^2$ and $Q_2^2$, respectively. 
The distributions at the final scales, $Q_1^2$ and $Q_2^2$, are constructed through the parton-to-parton functions, $E$, which themselves obey DGLAP-type~\cite{DGLAP} evolution equations:
\begin{equation}
\label{Eevo}
Q^2 \frac{\partial}{\partial Q^2} E_i^j(x,Q_0^2,Q^2)=\frac{\alpha_s(Q^2)}{2\pi}
\int_x^1 \frac{du}{u} P_k^i(u) E_i^k(x/u,Q_0^2,Q^2)\,,
\end{equation}
with initial condition $E_i^j(x,Q_0^2,Q_0^2)=\delta_i^j \delta(1-x)$ and  
$P_k^i(u)$ the Altarelli-Parisi splitting functions. The functions $E$ provide the 
resummation of collinear logarithms up to the accuracy with which the  $P_k^i(u)$ are specified.
We may therefore express $D_h^{j_1,j_2}(x_1,Q_1^2,x_2,Q_2^2)$ as
\begin{multline}
\label{Ddef}
D_h^{j_1,j_2}(x_1,Q_1^2,x_2,Q_2^2)=
\int_{x_1}^{1-x_2} \frac{dz_1}{z_1} \int_{x_2}^{1-z_1} \frac{dz_2}{z_2}  
\Big[ \\ D_h^{j_1',j_2'}(z_1,Q_0^2,z_2,Q_0^2) 
E_{j_1'}^{j_1}\Big( \frac{x_1}{z_1},Q_0^2,Q_1^2\Big) 
E_{j_2'}^{j_2}\Big( \frac{x_2}{z_2},Q_0^2,Q_2^2\Big) + \\
\int^{\mbox{\small{Min}}(Q_1^2,Q_2^2)}_{Q_0^2} d\mu_s^2  
D_{h,corr}^{j_1',j_2'}(z_1,z_2,\mu_s^2) 
E_{j_1'}^{j_1}\Big( \frac{x_1}{z_1},\mu_s^2,Q_1^2\Big) 
E_{j_2'}^{j_2}\Big( \frac{x_2}{z_2},\mu_s^2,Q_2^2\Big) \Big]\,. 
\end{multline}
The first term on r.h.s., usually addressed as the homogeneous term,
takes into account the uncorrelated evolution of the active partons found at a scale $Q_0^2$ in $D_h^{j_1',j_2'}$ up to $Q_1^2$ and $Q_2^2$, respectively. 
The second term, the so-called inhomogeneous one, takes into account the probability to find the active partons at $Q_1^2$ and $Q_2^2$ as a result of a splitting at a scale $\mu_s^2$, integrated over all the intermediate scale at which such splitting may occur. The distribution $D_{h,corr}^{j_1',j_2'}$ is given by
\begin{equation}
\label{Dcorr}
D_{h,corr}^{j_1',j_2'}(z_1,z_2,\mu_s^2)=
\frac{\alpha_s(\mu_s^2)}{2\pi \mu_s^2}
\frac{F_h^{j'}(z_1+z_2,\mu_s^2)}{z_1+z_2} 
\widehat{P}_{j'}^{j_1',j_2'} \Big( \frac{z_1}{z_1+z_2} \Big)\,.
\end{equation}
In eq.~(\ref{Dcorr}), $F_h^{j'}$  are single parton distributions 
and $\widehat{P}_{j'}^{j_1',j_2'}$ are the real
Altarelli-Parisi splitting functions~\cite{KUV}.
Both terms in eq.~(\ref{Ddef}) are shown in Fig.~(\ref{fig1}).
\noindent
The scale $Q_0^2$ is in general the (low) scale at which lDPDs are usually modelled, in complete analogy with the single-parton distributions case. In the present context it also acts as the factorisation scale for the correlated term, since all unresolved splittings, for which $\mu_s^2<Q_0^2$, are effectively taken into account in the parametrisation of $D_h^{j_1',j_2'}(z_1,Q_0^2,z_2,Q_0^2)$. In the ``equal scales'' case, taking the logarithmic 
derivative with respect to $Q^2$ in eq.~(\ref{Ddef}), we recover the result presented in 
Ref.~\cite{snigirev2003}:
\begin{multline}
\label{snigirev_LL}
Q^2\frac{\partial D_h^{j_1,j_2}(x_1,x_2,Q^2)}{\partial Q^2}=
\frac{\alpha_s(Q^2)}{2\pi}
\int_{\frac{x_1}{1-x_2}}^{1}\frac{du}{u} 
P_{k}^{j_1}(u) D_h^{j_2,k}(x_1/u,x_2,Q^2) + \\
\frac{\alpha_s(Q^2)}{2\pi}
\int_{\frac{x_2}{1-x_1}}^{1}\frac{du}{u} 
P_{k}^{j_2}(u) D_h^{j_1,k}(x_1,x_2/u,Q^2) +
\frac{\alpha_s(Q^2)}{2\pi}
\frac{F_h^{j'}(x_1+x_2,Q^2)}{x_1+x_2} \widehat{P}_{j'}^{j_1,j_2}
\Big( \frac{x_1}{x_1+x_2} \Big)\,.
\end{multline}
The first and second terms on the right-hand side are obtained through 
the $Q^2$ dependence contained in the $E$ functions, while the last 
is obtained from the $Q^2$ dependent limit in the $\mu_s^2$ integration
in the correlated term. The lDPDs evolution equations therefore resum 
large contributions of the type $\alpha_s \ln (Q^2/Q_0^2)$ and 
$\alpha_s \ln (Q^2/\mu_s^2)$ appearing in the uncorrelated and correlated term 
of eq.~(\ref{Ddef}), respectively.

\section{Momentum Sum Rule}
\label{MSR}
A number of sum rules for lDPDs has been already discussed and used to constrain
initial conditions for lDPDs evolution in Ref.~\cite{GS09}.
Sum rules are in general expected to hold on the basis of unitarity of the relevant
cross sections~\cite{DeTar}. In the following we show that the momentum sum rule for DPDs satisfies the necessary, but not sufficient for it to hold, condition of being 
preserved under QCD evolution. For this purpose we assume that the momentum sum rule is valid at an arbitrary but still perturbative scale $Q_0^2<Q_1^2,Q_2^2$:
\begin{equation}
\label{momentum_sum_rule0}
\sum_{j'_1} \int_0^{1-z_2} dz_1 z_1 D_h^{j'_1,j'_2}(z_1,Q_0^2,z_2,Q_0^2)=(1-z_2)F_h^{j'_2}(z_2,Q_0^2)\,,
\end{equation} 
which, as costumary, we choose as the starting scale for evolution. 
The aim of the calculation is therefore to verify that, once imposed at $Q_0^2$, DPDs fulfil the momentum sum rule 
\begin{equation}
\label{momentum_sum_rule}
\sum_{j_1} \int_0^{1-x_2} dx_1 x_1 D_h^{j_1,j_2}(x_1,Q_1^2,x_2,Q_2^2)=(1-x_2)F_h^{j_2}(x_2,Q_2^2)\,,
\end{equation} 
at any other scales $Q_1^2$ and $Q_2^2$, which we keep deliberately different. This strategy has been used to check an analogous sum rule in the context of dihadron fragmentation funtions~\cite{KUV} and fracture functions~\cite{Trentadue_Veneziano} while for lDPDs an explicit calculation has been presented in some detail in Ref.~\cite{BDFS_sum_rule}. We apply first the sum rule, eq.~(\ref{momentum_sum_rule}), to the homogeneous 
term in eq.~(\ref{Ddef})
\begin{equation}
\label{Ddef_hom}
\sum_{j_1} \int_0^{1-x_2} dx_1 x_1 
\int_{x_1}^{1-x_2} \frac{dz_1}{z_1} \int_{x_2}^{1-z_1} \frac{dz_2}{z_2} 
D_h^{j_1',j_2'}(z_1,Q_0^2,z_2,Q_0^2)  E_{j_1'}^{j_1}\Big( \frac{x_1}{z_1},Q_0^2,Q_1^2\Big) 
E_{j_2'}^{j_2}\Big( \frac{x_2}{z_2},Q_0^2,Q_2^2\Big)\,.
\end{equation}
We note that, reordering the $z_1$ and $x_1$ integrals,
the function $E_{j_1'}^{j_1}$ can be evaluted through the momentum sum rule for $E$-functions
which reads
\begin{equation}
\label{Esum}
\sum_{j_1} \int_0^1 dz z E_{j_1'}^{j_1}(z,Q_0^2,Q_1^2)=1\,.
\end{equation}
This property is derived in the appendix~\ref{P_sum_rules} and it might be thought as the parton level analogue of the momentum sum rule for fragmentation functions. 
Quite importantly, by using eq.~(\ref{Esum}), any explicit dependence on $Q_1^2$
desappears and we obtain
\begin{equation}
\label{Ddef_hom1}
\int_{0}^{1-x_2} dz_1 z_1\sum_{j_1'} \int_{x_2}^{1-z_1} \frac{dz_2}{z_2} 
D_h^{j_1',j_2'}(z_1,Q_0^2,z_2,Q_0^2) E_{j_2'}^{j_2}\Big( \frac{x_2}{z_2},Q_0^2,Q_2^2\Big)\,.
\end{equation}
Reordering again the integrals we get
\begin{equation}
\label{Ddef_hom2}
\int_{x_2}^{1} \frac{dz_2}{z_2} \Bigg[ \sum_{j_1'} \int_{0}^{1-z_2} dz_1 z_1  
D_h^{j_1',j_2'}(z_1,Q_0^2,z_2,Q_0^2) \Bigg] E_{j_2'}^{j_2}\Big( \frac{x_2}{z_2},Q_0^2,Q_2^2\Big) \,,
\end{equation}
where we now recognise in square brackets the sum rule at $Q_0^2$ in eq.~(\ref{momentum_sum_rule0}). 
For the homogeneous term we therefore obtain the following contribution
\begin{equation}
\label{Ddef_hom3}
\int_{x_2}^{1} \frac{dz_2}{z_2} (1-z_2)  
F_h^{j_2'}(z_2,Q_0^2) E_{j_2'}^{j_2}\Big( \frac{x_2}{z_2},Q_0^2,Q_2^2\Big)
\end{equation}
and conclude that its contribution alone does not reconstruct the expected result in eq.~(\ref{momentum_sum_rule}). We now turn to the inhomogeneous term in eq.~(\ref{Ddef}). 
Applying eq.~(\ref{momentum_sum_rule}) we get
\begin{multline}
\label{Ddef_inh}
\sum_{j_1} \int_0^{1-x_2} dx_1 x_1 
\int_{x_1}^{1-x_2} \frac{dz_1}{z_1} \int_{x_2}^{1-z_1} \frac{dz_2}{z_2} 
\cdot \\
\cdot \int^{Q_M^2}_{Q_0^2} d\mu_s^2 \frac{\alpha_s(\mu_s^2)}{2\pi \mu_s^2}
\frac{F_h^{j'}(z_1+z_2,\mu_s^2)}{z_1+z_2} 
\widehat{P}_{j'}^{j_1',j_2'} \Big( \frac{z_1}{z_1+z_2} \Big)
E_{j_1'}^{j_1}\Big( \frac{x_1}{z_1},\mu_s^2,Q_1^2\Big) 
E_{j_2'}^{j_2}\Big( \frac{x_2}{z_2},\mu_s^2,Q_2^2\Big)\,. 
\end{multline}
As in the previous case, the $E$-function which describes the evolution 
of the first parton can be integrated out by using $E$-momentum sum rule,
eq.~(\ref{Esum}). Changing then variable to $u=z_1/(z_1+z_2)$ we obtain
\begin{equation}
\label{sum_inh1}
\int^{Q_M^2}_{Q_0^2} d\mu_s^2 \frac{\alpha_s(\mu_s^2)}{2\pi \mu_s^2}
\int_{x_2}^{1} \frac{dz_2}{z_2} \int_{z_2}^{1} \frac{du}{u} \frac{1-u}{u}
F_h^{j'}\Big(\frac{z_2}{u},\mu_s^2\Big) \sum_{j'_1}
\widehat{P}_{j'}^{j_1',j_2'} (1-u) 
E_{j_2'}^{j_2}\Big( \frac{x_2}{z_2},\mu_s^2,Q_2^2\Big)\,.
\end{equation}
In order to proceed we need to relate real and regularised splitting functions. 
A number of these relations, valid in Mellin moment space, have been worked out
in Ref.~\cite{KUV}. The following relation is needed for our purpose
\begin{equation}
\label{real_to_reg:P}
\int_x^1 du (1-u) g(u) \sum_{j_1} \widehat{P}_{j'}^{j_1,j_2}(1-u) = \int_x^1 du (1-u) g(u) 
P_{j'}^{j_2}(u)\,,
\end{equation}
where $g(u)$ is a regular function of $u$. 
Since real and regularised splitting functions differ by the inclusion of virtual terms, proportional to $\delta(1-u)$ in the latter, the relation holds only at the integral level.
It generalises the symmetry of real splitting functions upon the exchange of daughter partons 
\begin{equation}
\label{Phat_symm}
\widehat{P}_{j'}^{j_1,j_2}(1-u)=\widehat{P}_{j'}^{j_2,j_1}(u)\,.
\end{equation}
Since this relation is not often encountered
in the literature we provide its explicit evaluation in appendix~\ref{real_to_reg}.
With the help of eq.~(\ref{real_to_reg:P}) and eq.~(\ref{Phat_symm}) we obtain 
\begin{equation}
\label{sum_inh2}
\int^{Q_M^2}_{Q_0^2} d\mu_s^2 \frac{\alpha_s(\mu_s^2)}{2\pi \mu_s^2}
\int_{x_2}^{1} \frac{dz_2}{z_2} 
\Bigg[ \int_{z_2}^{1} \frac{du}{u^2} - \int_{z_2}^{1} \frac{du}{u}\Bigg] 
F_h^{j'}\Big(\frac{z_2}{u},\mu_s^2\Big) 
P_{j'}^{j_2'} (u) E_{j_2'}^{j_2}\Big( \frac{x_2}{z_2},\mu_s^2,Q_2^2\Big)\,, 
\end{equation}
where we have split the $u$-integral in square brackets in two terms, A and B, respectively.
In order to deal with the $\mu_s^2$-integral, 
we need to build up a full derivative term with respect to $\mu_s^2$ out of the 
two $F\otimes P \otimes E$ terms in eq.~(\ref{sum_inh2}). For this purpose, we write $ F_h^{j'}$ as a solution of its evolution equation
\begin{equation}
\label{Fevo}
F_h^{j'} \Big( \frac{z_2}{u},\mu_s^2 \Big)=\int_{\frac{z_2}{u}}^{1} \frac{dw}{w}
F_h^r(w,Q_0^2) E_r^{j'}\Big( \frac{z_2}{uw},Q_0^2,\mu_s^2 \Big)\,,
\end{equation}
and substitute it in eq.~(\ref{sum_inh2}). For the B-term we get
\begin{multline}
\label{Bsum_inh}
B=-\int^{Q_M^2}_{Q_0^2} d\mu_s^2 \frac{\alpha_s(\mu_s^2)}{2\pi \mu_s^2}
\int_{x_2}^{1} \frac{dz_2}{z_2} 
\int_{z_2}^{1}\frac{du}{u}\int_{\frac{z_2}{u}}^{1} \frac{dw}{w}
F_h^r(w,Q_0^2) E_r^{j'}\Big( \frac{z_2}{uw},Q_0^2,\mu_s^2 \Big) \cdot \\ \cdot
P_{j'}^{j_2'} (u) E_{j_2'}^{j_2}\Big( \frac{x_2}{z_2},\mu_s^2,Q_2^2\Big)\,. 
\end{multline}
The $E_r^{j'} P_{j'}^{j_2'}$ term is a convolution over the $u$-variable
and has the basic structure of the r.h.s. of eq.~(\ref{Eevo}), so 
it can rewritten as a $\mu_s^2$-derivative:
\begin{equation}
\label{Bsum_inh2}
B=-\int^{Q_M^2}_{Q_0^2} d\mu_s^2 
\int_{x_2}^{1} dz_2  \int_{z_2}^{1} \frac{dw}{w}
F_h^r(w,Q_0^2) \Big[ \frac{\partial}{\partial \mu_s^2} 
E_r^{j'_2}\Big( \frac{z_2}{w},Q_0^2,\mu_s^2 \Big) \Big]
E_{j_2'}^{j_2}\Big( \frac{x_2}{z_2},\mu_s^2,Q_2^2\Big)\,. 
\end{equation}
Focusing now on the A-term, we would like to proceed as in the previous case 
and build a $\mu_s$-derivative out of the term $P_{j'}^{j_2'} E_{j_2'}^{j_2}$. 
In this case, however, the convolution variable and the matrix structure of the 
product do not allow a direct use of eq.~(\ref{Eevo}). In order to bring 
this term in a more manageable form, we first reorder the integrals 
and then change variables to the new convolution variable $y=x_2u/z_2$. We get
\begin{multline}
\label{Asum_inh}
A=\int^{Q_M^2}_{Q_0^2} d\mu_s^2 \frac{\alpha_s(\mu_s^2)}{2\pi \mu_s^2}
\int_{x_2}^{1} \frac{dw}{w} F_h^r(w,Q_0^2)
\int_{x_2/w}^{1} \frac{dy}{y^2} x_2 E_r^{j'}\Big( \frac{x_2}{yw},Q_0^2,\mu_s^2 \Big)
\cdot \\ \cdot
\int_{y}^{1}\frac{du}{u} P_{j'}^{j_2'} (u)
E_{j_2'}^{j_2}\Big( \frac{y}{u},\mu_s^2,Q_2^2\Big)\,.
\end{multline}
We notice that now the matrix structure of the term $P_{j'}^{j_2'}E_{j_2'}^{j_2}$ does correspond to the right-hand side of eq.~(\ref{Eevo}) transposed. Taking this into account, we may rewrite it as a $\mu_s^2$-derivative:
\begin{multline}
\label{Asum_inh2}
A=\int^{Q_M^2}_{Q_0^2} d\mu_s^2 
\int_{x_2}^{1} \frac{dw}{w} F_h^r(w,Q_0^2)
\int_{x_2/w}^{1} \frac{dy}{y^2} x_2 E_r^{j'}\Big( \frac{x_2}{yw},Q_0^2,\mu_s^2 \Big)
\Big[ - \frac{\partial}{\partial \mu_s^2} E_{j'}^{j_2}\Big( y,\mu_s^2,Q_2^2\Big)\Big]\,,
\end{multline}
where the extra minus sign comes from having differentiated the $E$-function with respect
to the lower scale. If now we change back to $z_2=x_2/y$, the A and B term 
have the same integral structure and can be summed together giving
\begin{multline}
A+B=-\int^{Q_M^2}_{Q_0^2} d\mu_s^2 
\int_{x_2}^{1} \frac{dw}{w} F_h^r(w,Q_0^2)
\int_{x_2}^{w} dz_2 \frac{\partial}{\partial \mu_s^2}\Big[
E_r^{j'}\Big( \frac{z_2}{w},Q_0^2,\mu_s^2 \Big)
E_{j'}^{j_2}\Big( \frac{x_2}{z_2},\mu_s^2,Q_2^2\Big) \Big]\,.
\end{multline}
Now the $\mu_s^2$ can be performed trivially and by using the initial conditions for the 
$E$-functions we obtain
\begin{multline}
A+B=- \int_{x_2}^{1} \frac{dw}{w} F_h^r(w,Q_0^2)
\int_{x_2}^{w} dz_2 \\ \Big[
E_r^{j'}\Big( \frac{z_2}{w},Q_0^2,Q_2^2 \Big) \delta_{j'}^{j_2}
\delta\Big(1-\frac{x_2}{z_2} \Big)
- \delta_{r}^{j'} \delta \Big( 1- \frac{z_2}{w} \Big) 
E_{j'}^{j_2}\Big( \frac{x_2}{z_2},Q_0^2,Q_2^2\Big) \Big]\,.
\end{multline}
Simplifying and changing $w$ to $z_2$ we get
\begin{equation}
A+B=\int_{x_2}^{1} \frac{dz_2}{z_2} F_h^r(z_2,Q_0^2)
(z_2-x_2) E_r^{j_2}\Big( \frac{x_2}{z_2},Q_0^2,Q_2^2 \Big)\,.
\end{equation}
Adding the result coming from the homogeneous calculation, eq.~(\ref{Ddef_hom3}),
we get 
\begin{equation}
(1-x_2) \int_{x_2}^{1} \frac{dz_2}{z_2} F_h^r(z_2,Q_0^2)
 E_r^{j_2}\Big( \frac{x_2}{z_2},Q_0^2,Q_2^2 \Big)\,.
\end{equation}
By using eq.~(\ref{Fevo}) we may rewrite the above convolution simply as $F_h^{j_2}(x_2,Q_2^2)$ and finally obtain the desidered momentum sum rule in eq.~(\ref{momentum_sum_rule}).
The large number of convolution integrals involved renders however 
such calculation not really transparent.
There is, however, an easier way to obtain the same result, that is to apply 
eq.~(\ref{momentum_sum_rule}) directly to the lDPDs evolution equations, eq.~(\ref{snigirev_LL}), and verify that, with this procedure, one recovers single PDFs evolution. 
We will present this calculation in the ``equal scales'' case.
When the sum rules operates on the left hand side of eq.~(\ref{snigirev_LL}), it simply gives the scale derivative of ordinary parton distributions weighted by the factor $(1-x_2)$:
\begin{equation}
\label{eq1}
(1-x_2)Q^2 \frac{\partial}{\partial Q^2} F_h^{j_2}(x_2,Q^2).
\end{equation}
Applying the momentum sum rule
the first term on the right hand side of eq.~(\ref{snigirev_LL}), which corresponds to the uncorrelated evolution of the first parton, we obtain 
\begin{equation}
\label{term:A1}
\frac{\alpha_s(Q^2)}{2\pi} \sum_{j_1} \int_0^{1-x_2} dx_1 x_1
\int_{\frac{x_1}{1-x_2}}^{1}\frac{du}{u} 
P_{k}^{j_1}(u) D_h^{j_2,k}(x_1/u,x_2,Q^2)\,.
\end{equation} 
Reordering the integrals and changing variable to $y=x_1/u$ we get
\begin{equation}
\label{term:A2}
\frac{\alpha_s(Q^2)}{2\pi} \sum_{j_1} 
\int_0^1 du u P_{k}^{j_1}(u) \int_0^{1-x_2}
dy y D_h^{k,j_2}(y,x_2,Q^2)\,.
\end{equation}
Exploiting now the basic property of splitting functions 
\begin{equation}
\label{P_sumrule}
\sum_{j_1} \int_0^1 du u P_{k}^{j_1}(u)=0\,,
\end{equation}
the term corresponding to the evolution of parton first vanishes, \textsl{i.e.} the overall momentum carried by the first parton is a quantity conserved by evolution. 
We now apply the sum rule to the second term of eq.~(\ref{snigirev_LL}), which corresponds to the uncorrelated evolution of the second parton and obtain 
\begin{equation}
\frac{\alpha_s(Q^2)}{2\pi} \sum_{j_1} \int_0^{1-x_2} dx_1 x_1
\int_{\frac{x_2}{1-x_1}}^{1}\frac{du}{u} 
P_{k}^{j_2}(u) D_h^{j_1,k}(x_1,x_2/u,Q^2)\,.
\end{equation}
Reordering the integrals we obtain
\begin{equation}
\label{eq4}
\frac{\alpha_s(Q^2)}{2\pi} 
\int_{x_2}^{1} \frac{du}{u} P_{k}^{j_2}(u)
\Bigg[ \sum_{j_1} \int_0^{1-x_2/u} dx_1 x_1 D_h^{j_1,k}(x_1,x_2/u,Q^2) \Bigg]\,.
\end{equation}
We recognise in square brackets the momentum sum rule written 
for momentum fraction $x_1$ and $x_2/u$ so that eq.~(\ref{eq4}) becomes
\begin{equation}
\label{eq2}
\frac{\alpha_s(Q^2)}{2\pi} 
\int_{x_2}^{1} \frac{du}{u} P_{k}^{j_2}(u)
\Big[1 -\frac{x_2}{u} \Big] F_h^k\Big(\frac{x_2}{u},Q^2\Big)\,.
\end{equation}
We interpret the term in square brackets as the fractional momentum of the proton $(1)$
minus the fractional momentum of the second parton before evolution, $x_2/u$.
We finally handle the correlated term in the evolution equations,
\begin{equation}
\sum_{j_1} \int_0^{1-x_2} dx_1 x_1
\frac{\alpha_s(Q^2)}{2\pi}
\frac{F_h^{j'}(x_1+x_2,Q^2)}{x_1+x_2} \widehat{P}_{j'}^{j_1,j_2}
\Big( \frac{x_1}{x_1+x_2} \Big)\,.
\end{equation}
We change variables to $u=x_2/(x_1+x_2)$ and get
\begin{equation}
\frac{\alpha_s(Q^2)}{2\pi} 
\int_{x_2}^{1} \frac{du}{u} x_2 \frac{1-u}{u} 
F_h^{j'}\Big(\frac{x_2}{u},Q^2\Big) \sum_{j_1} \widehat{P}_{j'}^{j_1,j_2}(1-u)\,.
\end{equation}
With the help of eq.~(\ref{real_to_reg:P}) and eq.~(\ref{Phat_symm}), we obtain
\begin{equation}
\label{eq5}
\frac{\alpha_s(Q^2)}{2\pi} 
\int_{x_2}^{1} \frac{du}{u} \Big[\frac{x_2}{u}-x_2\Big] 
F_h^{j'}\Big(\frac{x_2}{u},Q^2\Big) P_{j'}^{j_2}(u)\,.
\end{equation}
Again it is interesting to interpret the result and to give an explanation 
to the factor $x_2/u-x_2$. The momentum fraction of the (second) interacting
parton prior to the branching is $x_2/u$ while after the branching it has 
a fixed momentum fraction $x_2$. Therefore the quantity in square brackets is simply the 
fractional momentum of the first, integrated-over, parton.
If we now sum the results in eq.~(\ref{eq2}) and eq.~(\ref{eq4}), the
$u$-dependent terms cancel each other, and the net result is 
\begin{equation}
\frac{\alpha_s(Q^2)}{2\pi} \int_{x_2}^{1} \frac{du}{u} P_{k}^{j_2}(u) \Big[1-x_2\Big] 
F_h^k\Big(\frac{x_2}{u},Q^2\Big)\,.
\end{equation}
Equating this result to eq.~(\ref{eq1}), the factor $1-x_2$ can be simplified on both side and we obtain the familiar sPDF evolution equation for the second parton~\cite{DGLAP}: 
\begin{equation}
Q^2 \frac{\partial}{\partial Q^2} F_h^{j_2}(x_2,Q^2) =
\frac{\alpha_s(Q^2)}{2\pi}
\int_{x_2}^{1} \frac{du}{u} P_{k}^{j_2}(u)
F_h^k\Big(\frac{x_2}{u},Q^2\Big)\,. 
\end{equation}
From both calculations, it is clear the crucial role played by the inhomogeneous
term in order that the momentum sum rule for DPDs is preserved under QCD evolution. When the latter is directly applied to it, it takes into account the amount of fractional momentum lost by the second parton due to perturbative parton emissions. Or in other words, the contributions to the momentum sum 
rule coming from initial state radiation.
If such term is neglected altogether, the consistency of the formalism is lost since 
longitudinal momentum correlations are not properly taken into account. 
For these reasons, checking the sum rule is a useful method
to investigate the consistency of the evolution equations.

\section{Kinematic of the splitting term}
\label{kine}
\noindent
In a previous paper~\cite{Ceccopieri_DPD} we have proposed the DPDs evolution equations at different virtualities. This case is potentially relevant since many experimental 
DPS analyses consider the associated production of an electroweak boson, $Q_2^2\simeq M_{W^{\pm},Z}^2$, with jets, $Q_1^2 \simeq P_T^2$, where $P_T$ is the jet transverse momentum
typically choosen to be larger than $15$ GeV at LHC. Since DPS contributions 
are expected to populate low-$p_t$ particle spectrum, one may trigger on 
identified particles rather than on jets. This case, in which again $Q_2^2\gg Q_1^2$, has been considered in Ref.~\cite{LHCb_ZD} for the associated production of a $Z$-boson and 
a $D$-meson in the forward region of pp collisions at LHC and whose SPS background can be evaluated by using the results presented in Refs.~\cite{aDY_ceccopieri}. 
Turning back to the DPDs evolution equations at different virtualities and, to be definite,  considering the case $Q_1^2<Q_2^2$, we have found that the proposed homogeneous evolution equations with respect to the higher scale $Q_2^2$ does not fulfil the momentum sum rule,
eq.~(\ref{momentum_sum_rule}).
The disappearance of the inhomogeneous term was caused by the choice of the upper limit of the $\mu_s^2$ integral which, from strong ordering of virtualities implied by leading logarithmic approximation, was choosen to be $\mbox{Min}(Q_1^2,Q_2^2)$, and therefore independent of $Q_2^2$. This fact has induced us to reconsider in more detail the kinematics 
of the branching at the basis of the inhomogeneous term.
We first note that a physically plausible scale characterising the parton branching in the inhomogeneous term could be identified with the relative 
transverse momentum squared, $\bm{r}_T^2$, between the daughter partons, rather than the generic $\mu_s^2$ scale. This scale choice ambiguity can be resolved only within 
a higher order calculation. We can easily calculate this quantity considering a generic branching $p_0(1,\bm{0}_T,p_0^2)\rightarrow p_1(z,\bm{r}_T,p_1^2)+p_2(1-z,-\bm{r}_T,p_2^2)$, where we have explictely indicated longitudinal momentum fractions, transverse momenta and space-like virtualities of the relevant partons.
By performing a Sudakov decomposition of the four-momenta and setting $p_i^2=-k_i^2$
with $k_i^2>0$, it is then easy to show (see Appendix~\ref{deco}) that 
\begin{equation}
\label{rt_branching}
\bm{r}_T^2=(1-z)k_1^2+zk_2^2-z(1-z)k_0^2\,,
\end{equation}
where, in our notation, the fractional momentum $z$ is simply given by $z=z_1/(z_1+z_2)$.
Since the two floating scales $k_1^2$ and $k_2^2$ can
take values up to $Q_1^2$ and $Q_2^2$, the maximum value of the relative transverse
momentum at each branching is a function both of $Q_1^2$ and $Q_2^2$. 
Within a leading logarithmic approximation we can set this value to be 
\begin{equation}
Q_M^2=\epsilon_1 Q_1^2 + \epsilon_2 Q_2^2\,,
\end{equation}
with $\epsilon_1$ and $\epsilon_2$ being arbitrary constants of order one. 
This change, which is irrelevant in a leading logarithmic approximation,
induces however a $Q_2^2$ dependence in the upper integration limit of the 
$\bm{r}_T^2$-integral so that the inhomogeneous term in eq.~(\ref{Ddef}) now reads
\begin{multline}
\label{Dinh}
\int_{x_1}^{1-x_2} \frac{dz_1}{z_1} \int_{x_2}^{1-z_1} \frac{dz_2}{z_2}  
\int^{Q_M^2}_{Q_0^2} d\bm{r}_T^2
D_{h,corr}^{j_1',j_2'}(z_1,z_2,\bm{r}_T^2) 
E_{j_1'}^{j_1}\Big( \frac{x_1}{z_1},\bm{r}_T^2,Q_1^2\Big) 
E_{j_2'}^{j_2}\Big( \frac{x_2}{z_2},\bm{r}_T^2,Q_2^2\Big) \Big]\,. 
\end{multline}
This dependence implies that the resulting DPDs evolution equations will again contain an inhomogeneous term. Since the derivation is analogous to the one 
presented in Ref.~\cite{Ceccopieri_DPD} we just quote the result:
\begin{multline}
\label{DPDevo_Q2geQ1}
Q_2^2\frac{\partial D_h^{j_1,j_2}(x_1,Q_1^2,x_2,Q_2^2)}{\partial Q_2^2}=
\frac{\alpha_s(Q_2^2)}{2\pi}
\int_{\frac{x_2}{1-x_1}}^{1}\frac{du}{u} 
P_{k}^{j_2}(u) D_h^{j_1,k}(x_1,Q_1^2,x_2/u,Q_2^2) + \\
\frac{\alpha_s(Q_2^2)}{2\pi}
\frac{F_h^{j'}(x_1+x_2,Q_2^2)}{x_1+x_2} \widehat{P}_{j'}^{j_1,j_2}
\Big( \frac{x_1}{x_1+x_2} \Big)\,,
\end{multline}
where the initial conditions to the above evolution equations are the DPDs at 
$Q_0^2$ evolved up to $Q^2=Q_1^2$ with the usual ``equal scales'' 
evolution equations, eq.~(\ref{snigirev_LL}). It is then easy to show that 
eq.~(\ref{DPDevo_Q2geQ1}) satisfies the momentum sum rule, eq.~(\ref{momentum_sum_rule}).
Given the relation between relative transverse momentum and virtualities
in eq.~(\ref{rt_branching}), the appearance of the inhomogeneous term 
in eq.~(\ref{DPDevo_Q2geQ1}) can further justified noting that it takes into account 
the possibility that an asymmetric configuration of virtualities could be generated 
in a single parton branching in the last step of the evolution. 
As a last remark, it is interesting to note that 
setting $Q_1^2=Q_2^2=Q^2$ in eq.~(\ref{Dinh}) it is then possible to identify
with the integrand of eq.~(\ref{Dinh}) the $\bm{r}_T^2$-unintegrated version of 
DPDs
\begin{multline}
\label{Dinh2}
\mathcal{D}_h^{j_1,j_2}(x_1,x_2,Q^2,\bm{r}_T^2)=
\int_{x_1}^{1-x_2} \frac{dz_1}{z_1} \int_{x_2}^{1-z_1} \frac{dz_2}{z_2}  
D_{h,corr}^{j_1',j_2'}(z_1,z_2,\bm{r}_T^2) \cdot \\ \cdot 
E_{j_1'}^{j_1}\Big( \frac{x_1}{z_1},\bm{r}_T^2,Q^2\Big) 
E_{j_2'}^{j_2}\Big( \frac{x_2}{z_2},\bm{r}_T^2,Q^2\Big) \,,
\end{multline}
which is valid at fixed value of $\bm{r}_T^2$ in the range $Q_0^2<\bm{r}_T^2<Q^2$. 
Since in eq.~(\ref{Dinh2}) all the $Q^2$ dependences
are contained in the $E$-functions, it easy to show that the corresponding
evolution equations for the unintegrated $\mathcal{D}$ are homogeneous and
read
\begin{multline}
\label{Dbar_hom_rT}
Q^2\frac{\partial \mathcal{D}_h^{j_1,j_2}(x_1,x_2,Q^2,\bm{r}_T^2)}{\partial Q^2}=
\frac{\alpha_s(Q^2)}{2\pi}
\int_{\frac{x_1}{1-x_2}}^{1}\frac{du}{u} 
P_{k}^{j_1}(u) \mathcal{D}_h^{j_2,k}(x_1/u,x_2,Q^2,\bm{r}_T^2) + \\
+\frac{\alpha_s(Q^2)}{2\pi}
\int_{\frac{x_2}{1-x_1}}^{1}\frac{du}{u} 
P_{k}^{j_2}(u) \mathcal{D}_h^{j_1,k}(x_1,x_2/u,Q^2,\bm{r}_T^2)\,.
\end{multline}
It should be noted, however, that $\bm{r}_T^2$ is not observable, at variance
with the analogous case for extended dihadron~\cite{extDiff} and fracture functions
~\cite{extendedM}. 
Moreover $\mathcal{D}$ can not be readily interpreted 
as the distribution in relative transverse momentum of the interacting parton pair 
at the scale $Q^2$ since all transverse momentum generated during 
the evolution up to $Q^2$ is neglected by the $E$-functions, which 
are derived in the collinear approximation. For this interpretation 
to be correct one would need to replace the $E$-functions with appropriate 
Sudakov-like form factors~\cite{KT,AR}. Since, in general,  they tend
to broaden the transverse momemtum distribution as the final scale increases~\cite{CT,AR},
the distribution of the relative transverse momentum at $Q^2$ will have 
a broader tail with respect to the $\bm{r}_T^2$ distribution. 
Nevertheless this result has some formal resemblance
with the ones presented in Section 13 of Ref.~\cite{DS} and it remains 
for a future task to explore whether this fact is accidental or 
has deeper motivations.

\section{Summary}
\label{summary}
\noindent
We have presented two equivalent consistency checks for the momentum sum rule for DPDs and showed the importance of the inclusion of the so called 
inhomogeneous term in order to obtain these results. If such term is neglected altogether, the consistency of the formalism is lost since 
longitudinal momentum correlations are not properly taken into account.
Satisfying these consistency checks therefore impose strong constraint 
on the structure of DPDs evolution equations. With this respect we have 
revisited the result of Ref.~\cite{Ceccopieri_DPD} and, by a careful
rexamination of the kinametics of the splitting term, we have 
shown the DPDs evolution equations at different virtualities does contain
an inhomogeneous term.

\appendix
\section{}
\label{P_sum_rules}
\noindent
We report in this appendix the derivation of the sum rule for the second moment
of the $E$-function introduced in eq.~(\ref{Esum}). 
We first take the second moment of the $E$ evolution equations in eq.~(\ref{Eevo})
which then reads
\begin{equation}
\label{Eevo_1moment}
Q^2 \frac{\partial}{\partial Q^2} E_{i,2}^j(Q_0^2,Q^2)=\frac{\alpha_s(Q^2)}{2\pi}
 A_{k,2}^j E_{i,2}^k(Q_0^2,Q^2)\,,
\end{equation}
and where we have introduced the Mellin transforms
\begin{eqnarray}
E_{i,2}^k(Q_0^2,Q^2)&=&\int_0^1 dz z E_{i}^k(z,Q_0^2,Q^2)\,, \nonumber\\
A_{k,2}^j&=&\int_0^1 dz z P_k^j(z)\,.\nonumber
\end{eqnarray}
We may now sum eq.~(\ref{Eevo_1moment}) over the index $j$ 
\begin{equation}
Q^2 \frac{\partial}{\partial Q^2} \sum_j E_{i,2}^j(Q_0^2,Q^2)=\frac{\alpha_s(Q^2)}{2\pi}
 \sum_j A_{k,2}^j E_{i,2}^k(Q_0^2,Q^2)\,,
\end{equation}
and exploit the following property of anomalous dimensions (the moment space analogue
of eq.~(\ref{P_sumrule}) for splitting functions)
\begin{equation}
\label{SF_property}
\sum_j A_{k,2}^j=0\,,
\end{equation}
obtaining
\begin{equation}
Q^2 \frac{\partial}{\partial Q^2} \sum_j E_{i,2}^j(Q_0^2,Q^2)=0\,.
\end{equation}
This equation can be easily integrated to give
\begin{equation}
\sum_j E_{i,2}^j(Q_0^2,Q^2)-\sum_j E_{i,2}^j(Q_0^2,Q_0^2)=0\,.
\end{equation}
By exploiting the initial condition for the $E$-functions in moment space,
we get the desidered result
\begin{equation}
\label{Esum2}
\sum_j E_{i,2}^j(Q_0^2,Q^2)=1\,,
\end{equation}
valid for any fixed value of the index $i$. It appears therefore 
that the moment sum rule for the $E$-functions is a direct consequences
of eq.~(\ref{SF_property}) which, in turn, follows, for example, from 
conservation of overall quarks plus gluons momenta in a proton at any 
value of $Q^2$. 

\section{}
\label{real_to_reg}
\noindent
In this appendix we explicitly check the identity quoted in the text for the 
particular case $j'=q$ and $j_2=q$. With this settings we have
\begin{equation}
\label{real_to_reg:ex1}
\int_x^1 du (1-u) g(u) \widehat{P}_{q}^{g,q}(1-u) = \int_x^1 du (1-u) g(u) 
P_{q}^{q}(u)\,,
\end{equation}
where the sum over $j_1$ collapsed to just one term ($j_1=g$) due to the nature 
of vertex of leading order splitting functions. 
Substituting the relevant splitting functions 
\begin{equation}
\widehat{P}_{q}^{g,q}(u)=C_F \frac{1+(1-u)^2}{u}\,, \;\;\;
P_{q}^{q}(u)=C_F \Big( \frac{1+u^2}{1-u} \Big)_+\,.
\end{equation}
in eq.~(\ref{real_to_reg:ex1}), the latter reduces to
\begin{equation}
\label{real_to_reg:ex2}
\int_x^1 du g(u)  (1+u^2) = \int_x^1 du \frac{1+u^2}{1-u} [f(u)-f(1)]
-f(1) \int_0^x du \frac{1+u^2}{1-u}\,,  
\end{equation}
where we have introduced an auxiliary function $f(u)=(1-u) g(u)$ and exploited 
the standard definition of plus distribution. Since $g(u)$ is a regular 
function of $u$, it follows that $f(1)=0$ and the identity is easily proved.
For the other splitting function which involves virtual contributions, namely $P_g^g(u)$,  
the calculation is analogous. Therefore, for the purpose of restoring symmetry of splitting functions upon the exchange of the daughter partons $j_1$ and $j_2$, the weighting function $(1-u)$ is instrumental to let the virtual terms, proportional to $\delta(1-u)$ and at the origin of symmetry breaking, vanish.

\section{}
\label{deco}
In this appendix we report the calculation used to arrive at eq.~(\ref{rt_branching}).
See also Section 2.3 of Ref.~\cite{Gieseke}. We consider the parton branching $0(p_0)\rightarrow 1(p_1)+2(p_2)$ with four-momenta 
in parenthesis decomposed as follows
\begin{eqnarray}
\label{c1}
p_0^2&=&-k_0^2,\; k_0^2>0\,\nonumber\\
p_1&=&z p_0 + r_T + \xi_1 \eta,\\
p_2&=&(1-z) p_0 - r_T + \xi_2  \eta, \nonumber
\end{eqnarray}
where $\eta$ is a lightlike vector ($\eta^2=0$)
and $r_T=(0,\bm{r}_T,0)$ is the relative transverse momentum such that
$r_T^2=-\bm{r}_T^2$, $p_0 \cdot r_T=\eta \cdot r_T=0$ and $p_0 \cdot \eta \neq 0$\,. The two parameters $\xi_1$ and $\xi_2$ can be obtained 
by imposing on eqs.~(\ref{c1}) the mass-shell relations
\begin{eqnarray}
\label{c2}
p_1^2&=&-k_1^2,\; k_1^2>0\,,\nonumber\\
p_2^2&=&-k_2^2,\; k_2^2>0\,.
\end{eqnarray}
We obtain
\begin{equation}
\label{c3}
\xi_1=\frac{-k_1^2+z^2k_0^2-r_T^2}{2zp_0 \cdot \eta}, \;\;\; 
\xi_2=\frac{-k_2^2+(1-z)^2k_0^2-r_T^2}{2(1-z)p_0 \cdot \eta}\,.
\end{equation}
Squaring the momentum conservation equation, $p_0=p_1+p_2$, we get 
\begin{equation}
2 p_0 \cdot \eta (\xi_1+\xi_2)=0\,.
\end{equation}
Substituting the values for $\xi_i$ and after some algebra we arrive at the desiderd result 
\begin{equation}
\bm{r}_T^2=(1-z)k_1^2+zk_2^2-z(1-z)k_0^2\,.
\end{equation}


\begin{thebibliography}{99}
\bibitem{DPS_exp_new}G.~Aad \textit{et al.} (ATLAS Collaboration), \textsl{New J.~Phys.~} \textbf{15} (2013) 033038\,;\\
R.~Aaij \textit{et al.} (LHCb Collaboration), \textsl{JHEP} \textbf{06} (2012) 141\,;\\
F.~Abe \textit{et al.} (CDF Collaboration), \textsl{Phys.~Rev.~}  \textbf{D56} (1997) 3811\,;\\
V.~M.~Abazov  \textit{et al.} (D0 Collaboration) \textsl{Phys.~Rev.~} \textbf{D81} (2010)  052012\,;\\
V.~M.~Abazov  \textit{et al.} (D0 Collaboration) \textsl{Phys.~Rev.~} \textbf{D83} (2011) 052008 \,;\\
S.~Chatrchyan \textit{et al.} (CMS Collaboration), arXiv:1312.5729 [hep-ex]\,;\\
S.~Chatrchyan \textit{et al.} (CMS Collaboration), arXiv:1312.6440 [hep-ex]\,.
\bibitem{snigirev2003} A.~M.~Snigirev, \textsl{Phys.~Rev.~} \textbf{D68} (2003) 114012\,.
\bibitem{Ceccopieri_DPD} F.~A.~Ceccopieri, \textsl{Phys.~Lett.~} \textbf{B697} (2011) 482.
\bibitem{BDFS_sum_rule} B.~Blok \textit{et al.}, arXiv:1306.3763 [hep-ph]\,.
\bibitem{MW}A.~V.~Manohar and W.~J.~Waalewijn, \textsl{Phys.~Lett.~} \textbf{B713} (2012) 196\,;\\ \textsl{Phys.~Rev.~} \textbf{D85} (2012) 114009\,.
\bibitem{DS}M.~Diehl and A.~Schafer, \textsl{Phys.~Lett.~} \textbf{B698} (2011) 389\,;\\
M.~Diehl, D.~Ostermeier and A.~Schafer, \textsl{JHEP} \textbf{1203} (2012) 089\,.
\bibitem{D014}V.~M.~Abazov  \textit{et al.} (D0 Collaboration), arXiv:1402.1550 [hep-ex]\,.
\bibitem{DGLAP}L.N.~Lipatov,~\textsl{Sov.~J.~Nucl.~Phys.~}  \textbf{20} (1975) 95\,;\\
V.N.~Gribov and L.N.~Lipatov,~\textsl{Sov.~J.~Nucl.~Phys.~}  \textbf{15} (1972) 438\,;\\
G.~Altarelli and G.~Parisi,~\textsl{Nucl.~Phys.~} \textbf{B126} (1977) 298\,;\\
Yu.L.~Dokshitzer \textsl{Sov.~Phys.~JETP~} \textbf{46} (1977) 641.
\bibitem{BDFS_4jet} B.~Blok \textit{et al.}, \textsl{Phys.~Rev.~} \textbf{D83} (2011) 071501\,. 
\bibitem{BDFS_12} B.~Blok \textit{et al.}, \textsl{Eur.~Phys.~J.~} \textbf{C72} (2012) 1963\,.
\bibitem{GS11} J.~R.~Gaunt and W.~J.~Stirling, \textsl{JHEP} \textbf{1106} (2011) 048\,;\\
J.~R.~Gaunt, \textsl{JHEP} \textbf{1301} (2013) 042.
\bibitem{KUV}K.~Konishi, A.~Ukawa and G.~Veneziano, \textsl{Nucl.~Phys.~} \textbf{B157} (1979) 45.
\bibitem{GS09}J.~R.~Gaunt and W.~J.~Stirling, \textsl{JHEP} \textbf{1003} (2010) 005.
\bibitem{DeTar}C.~E.~DeTar, D.~Z.~Freedman and G.~Veneziano, \textsl{Phys.~Rev.~} \textbf{D4} (1971) 906\,.
\bibitem{Trentadue_Veneziano} L.~Trentadue and G.~Veneziano,~\textsl{Phys.~Lett.~} \textbf{B323} (1994) 201.
\bibitem{extendedM} G.~Camici, M.~Grazzini and L.~Trentadue,  \textsl{Phys.~Lett.~} \textbf{B439} (1998) 382.
\bibitem{LHCb_ZD}R.~Aaij \textit{et al.} (LHCb Collaboration), arXiv:1401.3245 [hep-ex]\,.
\bibitem{aDY_ceccopieri} F.~A.~Ceccopieri and L.~Trentadue, \textsl{Phys.~Lett.~} \textbf{B668} (2008) 319\,;\\
F.~A.~Ceccopieri, \textsl{Phys.~Lett.~} \textbf{B703} (2011) 491\,.
\bibitem{extDiff} F.~A.~Ceccopieri, M.~Radici and A.~Bacchetta, \textsl{Phys.~Lett.~} \textbf{B650} (2007) 81\,.
\bibitem{KT}J.~Kodaira and L.~Trentadue, \textsl{Phys.~Lett.~} \textbf{B112} (1982) 66\,;
\textsl{Phys.~Lett.~} \textbf{B123} (1983) 335\,.
\bibitem{AR}S.~M.~Aybat and T.~C.~Rogers, \textsl{Phys.~Rev.~} \textbf{D83} (2011) 114042\,. 
\bibitem{CT}F.~A.~Ceccopieri and L.~Trentadue, \textsl{Phys.~Lett.~} \textbf{B660} (2008) 43\,;\\ F.~A.~Ceccopieri and L.~Trentadue, \textsl{Phys.~Lett.~} \textbf{B636} (2006) 310\,.
\bibitem{Gieseke} S.~Gieseke, P.~Stephens, B.~Webber, \textsl{JHEP} \textbf{0312} (2003) 045\,. 
\end{thebibliography}
\end{document}